%Paper: hep-th/9212027
%From: mandal%theory%tifrvax.BITNET@CEARN.cern.ch
%Date: Fri, 4 Dec 92 01:51:40 IST

\input phyzzx

%%%%%%%%            PLEASE USE PHYZZX                   %%%%%%%%%%%%%
%%%%%%%%                                                %%%%%%%%%%%%%
%%%%%%%%   A  TIME-DEPENDENT CLASSICAL SOLUTION  OF     %%%%%%%%%%%%%
%%%%%%%%            C=1 STRING FIELD THEORY             %%%%%%%%%%%%%
%%%%%%%%         AND NON-PERTURBATIVE EFFECTS           %%%%%%%%%%%%%
%%%%%%%% AVINASH DHAR, GAUTAM MANDAL AND SPENTA R. WADIA%%%%%%%%%%%%%
%%%%%%%%                                                %%%%%%%%%%%%%
\def\winf{$W_\infty$}

\def\del{\partial}

\def\ni{\noindent}

\def\ket{\rangle}
\def\bra{\langle}

\def\D{{\cal D}}
\def\gst{{g_{\rm str}}}
\def\t{{\cal T}}
\def\rhot{{\tilde \rho}}

\indent \hfill{TIFR-TH-92/40}\break
\indent \hfill December, 1992 \break
\date{}
\titlepage
\title{\bf A Time-dependent Classical Solution of $c=1$ String Field Theory
and Non-perturbative Effects}
\author{Avinash Dhar\foot{adhar@tifrvax.bitnet.},
Gautam Mandal\foot{mandal@tifrvax.bitnet.}
and Spenta R. Wadia
\foot{wadia@tifrvax.bitnet}}
\address{ Tata Institute of Fundamental Research, Homi Bhabha Road, Bombay
400 005, India}
\abstract{We describe a real-time classical solution of $c=1$ string field
theory written in terms of the phase space density, $u(p,q,t)$, of the
equivalent fermion theory.  The solution corresponds to tunnelling of a
single fermion above the filled fermi sea and leads to amplitudes that go
as $\exp(- C/ \gst)$. We discuss how one can use this technique to
describe non-perturbative effects in the Marinari-Parisi model.  We also
discuss implications of this type of solution for the two-dimensional
black hole.}
\endpage

\REF\SHENKER{S. Shenker, in {\sl Random Surfaces and Quantum Gravity},
Cargese Proceedings, Eds. O. Alvarez, E. Marinari and P. Windey (Plenum,
1991).}
\REF\CONE{E. Brezin, V.A. Kazakov and Al.B. Zamolodchikov, Nucl. Phys.
338 (1990) 673; D.J. Gross and N. Milikovic, Nucl. Phys. B238 (1990) 217;
G. Parisi, Europhys Lett. 11 (1990) 595; P. Ginsparg and J. Zinn-Justin,
Phys. Lett. 240B (1990) 333; S.R. Das, A. Dhar, A. Sengupta and S.R.
Wadia, Mod. Phys. Lett. A5 (1990) 891.}
\REF\MP{E. Marinari and G. Parisi, Phys. Lett. 240B (1990) 375.}
\REF\DMWA{A. Dhar, G. Mandal and S.R. Wadia,
Mod. Phys. Lett. A7 (1992) 3129.}
\REF\BIPZ{E. Brezin, C. Itzyksen, G. Parisi and J.B. Zuber, Comm. Math.
Phys. 59 (1978) 35.}
\REF\DMWC{A. Dhar, G. Mandal and S.R. Wadia,
Tata preprint TIFR-TH-92/63, hep-th/9210120.}
\REF\DAS{S.R. Das, TIFR preprint, TIFR/TH/92-62.}
\REF\RUSSO{J.G. Russo, Texas preprint UTTG-27-92.}
\REF\MSW{G. Mandal, A.M. Sengupta and S.R. Wadia, Mod. Phys. Lett. A6
(1991) 1685.}
\REF\W{E. Witten, Phys. Rev. D44 (1991) 314.}
\REF\DDMW{S.R. Das, A. Dhar, G. Mandal and S.R. Wadia, Int. J. Mod. Phys.
A7 (1992) 5165.}
\REF\DMWB{ A. Dhar, G. Mandal and S.R. Wadia, preprint
IASSNS-HEP-91/89, TIFR/TH/91-61, hep-th/9204028
(to appear in Int. J. Mod. Phys.).}
\REF\POL{J. Polchinski, Nucl. Phys. B346 (1990) 253.}
\REF\SAKITA{S. Iso, D. Karabali and B. Sakita,
preprint CCNY-HEP-92/6.}
\REF\SW{
A.M. Sengupta and S.R. Wadia, Int. J. Mod. Phys. A6 (1991) 1961.}
\REF\DJ{
A. Jevicki and
B. Sakita, Nucl. Phys. B165 (1980) 511; S.R. Das and A.
Jevicki, Mod. Phys. Lett. A5 (1990) 1639.}
\REF\OVRUT{R. Brustein and B. Ovrut, Penn. U. preprint UPR-524T.}
\REF\MENDE{J. Lee and P.F. Mende, preprint BROWN-HET-880.}
\REF\JEVICKI{A. Jevicki, Nucl. Phys. B376 (1992) 75.}
\REF\M{G. Bhanot, G. Mandal and O. Narayan, Phys. Lett. B251 (1990) 388;
F. David, Nucl. Phys. B348 (1991) 507.}
\REF\HAWKING{S. Hawking and J.M. Stewart, DAMTP, Cambridge preprint
PRINT-92-0362.}
\REF\DMWD{A. Dhar, G. Mandal and S. Wadia, TIFR preprint, in preparation.}

\section{\bf Introduction:}

If we call $\gst$ the coupling constant of string theory then we would
expect that, at weak coupling, non-perturbative effects would go as $
\exp(- C/\gst^2)$. However, Shenker
[\SHENKER]  has made the remarkable observation
that in closed string theories there can be non-perturbative effects which
go as $\exp(- C/\gst)$ and that this is a generic feature of string
theory. Such a non-perturbative behaviour can be argued for on the basis of
large orders of string perturbation theory and can be seen in the solutions of
string theory with $c<1$. Their existence in the $c=1$ model
(two-dimensional string theory) [\CONE]
and in the Marinari-Parisi model [\MP] is also
argued for in terms of `eigenvalue tunnelling'. For a detailed discussion
of these we refer the reader to [\SHENKER].

In this paper we discuss this phenomenon in the non-perturbative
formulation of string field theory at $c=1$ deveolped in
 [\DMWA]. This formulation is based on the
mapping of the $c=1$ matrix model onto a theory of free non-relativistic
fermions moving in one dimension in a potential
[\BIPZ, \CONE]. The central object in the formulation developed in [\DMWA]
is the fermion density operator,
$u(p,q,t)= \int dx \psi^\dagger(q-x/2) \psi(q+ x/2) \exp(-ipx)$,
whose expectation
value in any state is the fermion distribution function in phase space in
that state. The string field theory action of [\DMWA] is written in terms
of $u(p,q,t)$ and has a nontrivial dependence on $\gst\sim \hbar$.
Formally, the action can be written as an infinite series in $\gst$.  We
present a real-time classical solution of this theory which can be seen to
correspond to quantum tunnelling of a single fermion above the filled
fermi sea. This classical solution has a nontrivial dependence on $\gst$
and leads to amplitudes that go as $\exp(-C/\gst)$ rather than
$\exp(-C/\gst^2)$.   A comparison with collective field theory  shows that
our classical solution is not a classical solution of the collective field
theory even in the limit $\gst \to 0$. We explain the discrepancy.  Our
classical solution can be generalized to the Marinari-Parisi model and has
a possible application to supersymmetry breaking in that model.
Also, in the context of identification [\DMWC, \DAS, \RUSSO]
with black hole physics in two dimensions [\MSW, \W]
we find that the ``hyperbolic'' transform [\DMWC] of
our solution corresponds to a rather interesting time-dependent tachyon
solution in the black hole background.

The plan of the paper is as follows. In Sec. 2 we briefly review our
formulation of the $c=1$ string field theory and set up the notation.
In Sec. 3 we discuss an exact solution of the classical equations of
motion of the $c=1$ string field theory which describes a
single fermion tunnelling. In Sec. 4 we combine this solution with the
phase space density corresponding to the filled fermi sea of $N-1$ fermions
to get a time-dependent solution of the full theory and show that it leads
to amplitudes $\propto \exp(-C/\gst)$. In Sec. 5, we discuss
how this technique can be applied to find non-perturbative effects
in the Marinari-Parisi Model. In Sec. 6, we make a comparsion with
collective field theory. In Sec. 7, we discuss our solution in the black
hole context.

\section{\bf $c=1$ String Field Theory:}

We briefly review the non-perturbative formulation of the $c=1$ string
field theory [\DMWA].  As is well-known, this theory is exactly described
by non-relativistic fermions moving in a background hamiltonian
[\BIPZ, \CONE].  The double scaled field theory corresponds to the
hamiltonian $h(p,q) = {1\over2} (p^2 - q^2)$.  Since the fermion number is
held fixed, the basic excitations are described by the bilocal operator
$\phi(x,y,t) = \psi(x,t) \psi^\dagger(y,t)$ or equivalently its transform
$$
u(p,q,t) = \int^{+\infty}_{-\infty} dx~\psi^\dagger
\left(q - \hbar x/2, t\right)
e^{-ipx} \psi\left(q + \hbar x/2, t\right)
\eqn\oneone
$$
Here and in the following we have used the notation $\hbar$ for $\gst$ as
in [\DMWA].  The expectation value of this operator in a state is the
phase space fermion distribution function in that state.
Eqn. \oneone\ also has the
important property that given a ``classical function'' $f(p,q,t)$ in the
phase space, we have an operator in the fermion field theory
$$
{\cal O}_f = \int {dpdq  \over (2\pi)^2}
f(p,q,t) u(p,q,t) = {1\over 2\pi}\int dx~\psi^\dagger(x,t) \hat
f(\hat x,\hat p,t) \psi(x,t)
\eqn\onetwo
$$
where $\hat f(\hat x,\hat p,t)$ is the Weyl-ordered operator corresponding
to the classical function $f(p,q,t)$.  For example, vector fields
corresponding to the functions $f_{\alpha\beta} (p,q) =
e^{i(p\beta-q\alpha)}$ satisfy the classical algebra $\omega_\infty$ of
area-preserving diffeomorphisms.  The corresponding quantum operators in
the fermion field theory
$$
\tilde u(\alpha,\beta,t) = \int {dp~dq \over
(2\pi)^2} e^{i(p\beta-q\alpha)} u(p,q,t)
\eqn\onethree
$$
satisfy the $W_\infty$ algebra (a one-parameter deformation of
$\omega_\infty$) \foot{We have in our previous works also
used the notation $W(\alpha,\beta,t)$ for $\tilde u(\alpha,\beta,t)$.}
$$
[\tilde u(\alpha,\beta,t),\tilde u(\alpha',\beta',t)] =
{i\over \pi}  \sin {\hbar
\over 2} (\alpha\beta' - \beta\alpha') \tilde u(\alpha + \alpha',\beta +
\beta',t)
\eqn\onefour
$$
An exact boson representation of the fermion field theory that reflects
the $W_\infty$ symmetry can be acheived in terms of the 3-dim. field
$u(p,q,t)$, provided we impose the constraints that follow from its
microscopic definition
$$
\int {dp~dq \over 2\pi \hbar} u(p,q,t) = N
\eqn\onefive
$$
$$
\cos {\hbar \over 2} \left(\partial_q \partial_{p'} - \partial_{q'}
\partial_p\right) u(p,q,t) u(p',q',t)\bigg|_{{p' = p \atop q' = q}} = u
(p,q,t)
\eqn\onesix
$$
where $N$ is the total number of fermions.  Also the equation of motion
that follows from the definition \oneone\ is
$$
(\partial_t + p\partial_q + q\partial_p) u(p,q,t) = 0
\eqn\oneseven
$$
The constraints \onefive\ and
\onesix\  in fact specify a co-adjoint orbit of
$W_\infty$, and the classical action is constructed using the
method of Kirillov
$$
\eqalign{
S[u,h] = &\int ds~dt \int {dp~dq \over 2\pi\hbar} u(p,q,t,s)
\hbar^2\left\{\partial_s u(p,q,t,s), \partial_t u(p,q,t,s)\right\}_{MB}
\cr & + \int dt \int {dp~dq \over 2\pi\hbar} u(p,q,t) h(p,q).}
\eqn\oneeight
$$
where $\{~,~\}_{MB}$ is the Moyal bracket (for details see
[\DMWA]).

We wish to emphasize that the action \oneeight, together with a measure in
the functional integral that incorporates  the constraints \onefive\ and
\onesix\ can be {\bf derived}, starting from the fermion field theory, by
the standard procedure of time-slicing and inserting complete sets of
$W_\infty$-coherent states. Thus, one can derive the following identity
for the $n$-point function of the bilocal fermion operator $u(p,q,t)$:
$$ \bra \mu | T(u(p_1, q_1, t_1)\cdots u(p_n, q_n, t_n) )| \mu \ket
= \int \D u\, u(p_1, q_1, t_1)\cdots u(p_n, q_n, t_n) \exp({i\over \hbar}
S[u, h]) \eqn\onenine $$
where $S[u,h]$ is the action \oneeight\ and the measure $\D u$ includes
$\delta$-functions incorporating the constraints \onefive\ and \onesix.
The state $| \mu \ket$ on the left hand side refers to the fermi ground
state.

Let us now briefly indicate the classical limit of the string theory
$(\hbar \rightarrow 0)$.  In this limit the constraint \onesix\ implies
that $u(p,q,t)$ is a characteristic function of a region of phase space
specified by a boundary [\DDMW, \DMWB, \POL, \SAKITA].
For example the ground state corresponds
to the static solution $u(p,q) = \theta(\mu-h(p,q))$, $\mu \sim
-{1\over\hbar}$.  The massless excitation (tachyon)
[\SW, \DJ] corresponds to a curve
that is a small deviation from the fermi surface $
h(p,q) = {1\over2}(p^2-q^2) =\mu$.
\bigskip

\section{\bf Time-dependent Classical Solution For Single-fermion
Tunnelling:}

We shall now describe a time-dependent classical solution of the
$u(p,q,t)$ theory \onefive-\oneeight\ that describes the phenomenon
of {\sl quantum mechanical} tunnelling of a single fermion through the
potential barrier.

We wish to emphasize at the outset that {\bf an effect which is  genuinely
quantum mechanical in terms of a single fermion can be described
entirely by a classical solution of the $u(p,q,t)$ theory in
real time}. Such a phenomenon is not  unfamiliar: the classical
Euler-Lagrange equation of a Schr\"odinger  field theory is identical to
the Schr\"odinger equation of single-particle quantum mechanics. The fact
that the classical theory of $u(p,q,t)$ describes the single-particle
quantum mechanics exactly is indicated by the appearance of explicit
factors of $\hbar$ in the classical action and the constraints, as well as
by the fact that the action is derived from coadjoint orbit of \winf\
rather than $w_\infty$.  The latter is the group of canonical
transformations in the classical single-particle phase space whereas the
former is the group of unitary transformations in the single-particle
Hilbert space. This, indeed, is the main difference between our formalism
and standard collective field theory [\DJ]--- classical solutions of the
latter describe only classical motion of the fermions and do not
accomodate their quantum fluctuations. We shall see this difference
quantitatively in Sec.  6.

\underbar{Single-fermion wave-packet in phase space:}

We shall first describe a solution $u_1(p,q,t)$ of
\onefive-\oneeight\ with $N$ in \onefive\ put equal to one.
This  phase space
density corresponds to a single isolated fermion  tunnnelling
across the potential barrier. Later we will combine this solution  with a
stationary solution corresponding to a fermi sea built out of $N-1$
fermions  to construct a solution of the full $N$-particle system.

It is easy to verify that
$$u_1(p,q,t) = 2 \exp{-1\over \hbar}[(p\cosh t - q\sinh t-p_0)^2 +
(-p\sinh t + q\cosh t - q_0)^2] \eqn\twoone$$
satisfies the equation of motion
$$\del_t u_1(p,q,t) = \{ h, u_1\}_{\sevenrm MB}=-(p\del_q + q\del_p)u_1
\eqn\twotwo $$
and the constraints
$$\int {dpdq\over 2\pi\hbar} u_1(p,q) = 1  \eqn\twothree$$
$$ \cos {\hbar \over 2} (\del_q \del_{p'} - \del_{q'} \del_p )
[u_1(p,q) u_1(p',q')]_{p'=p, q'=q} = u_1(p,q) \eqn\twofour$$

It is clear that $u_1(p,q,t)$ is a configuration that describes
the phase space density of a single
fermion. In the next section we shall see that $u_1(p,q,t)$ corresponds
to the phase space density of a fermion in a minimum uncertainty
wavepacket (Eqn. (39)).
Note that the peak of the phase space density at time $t$ is
given by
$$ p\cosh t - q\sinh t - p_0 =0 = -p\sinh t + q\cosh t-q_0 \eqn\twofive$$
The above equations give the position of the peak at time $t$ as
$$ \bar p(t) = p_0\cosh t + q_0 \sinh t, \quad \bar q(t)= p_0 \sinh t +
q_0
\cosh t \eqn\twosix$$
Let us choose $p_0>0, q_0<0$ and $p_0 <|q_0|$ for definiteness, so that
the mean trajectory \twosix\ describes a hyperbola in the left half space
corresponding to negative energy (negative value of $(p_0^2-q_0^2)/2$).
We shall
equivalently use a parametrization
$$ p_0 = \sqrt{2|E_0|}\sinh \theta_0, \qquad q_0=-\sqrt{2|E_0|}
\cosh\theta_0
\eqn\twoseven $$
where $E_0= - |E_0|= (p_0^2 - q_0^2)/2$ denotes the energy of the
trajectory.

The trajectory of the peak, \twosix, suggests that classically the
fermion is  completely reflected off the barrier. To see this more
quantitatively, note that in the $\hbar \to 0$ limit we get
$$\eqalign{
{1\over 2\pi\hbar}
u_1(p,q,t) \to &\delta(p\cosh t - q\sinh t - p_0)\delta( -p\sinh t +
q\cosh t-q_0 )\cr
 =&\delta(p- \bar p(t))\delta(q- \bar q(t)) \cr}
\eqn\twoeight$$
where $\bar p(t)$ and $ \bar q(t)$ are given by \twosix.
For finite $\hbar$, however, the phase space density has a finite spread
and a finite amount of the phase space density trickles across to the
other side of the potential barrier. The easiest way to see this is to
look at the fermion density $\rho(q,t)$:
$$\eqalign{
\rho(q,t) \equiv & \int {dp\over 2\pi\hbar} u_1(p,q,t) \cr
=& (\pi\hbar \cosh 2t)^{-1/2}
\exp[-{ (q - \bar q(t))^2 \over \hbar \cosh 2t}] \cr}
\eqn\twonine$$
where $\bar q(t)$ has been defined in \twosix.

There are several interesting facts about \twonine. First of all, it is
defined for all $q$, positive and negative. Therefore, there is a non-zero
probability density of the fermion in the right half  of the world ($q>0$)
at any time. Moreover, although the mean position $\bar q(t)$ again shows
the classically reflected trajectory, the dispersion
$$\Delta q (t) = \sqrt {{\hbar\over 2} \cosh 2t}$$
increases exponentially rapidly at large times (positive as well as
negative). This means that the density \twonine\ is reasonably peaked at
finite times around its mean but at large negative  and positive times it
gets very spread out. How does one find out if there is a finite amount of
probability that actually moves over from the left side of the barrier to
the right?

Let us consider the total probability, at any given time, of
the fermion to be in the right (or left) half of $q$-space. In other
words, we define
$$N_+(t) = \int_0^\infty  dq\, \rho(q,t) \eqn\twoten$$
and
$$N_-(t) = \int_{-\infty}^0  dq\, \rho(q,t) \eqn\twotena$$
By \twothree, $N_+(t) + N_-(t)= 1$; hence only one of them is independent.
We shall focus on $N_+(t)$.
Using \twonine, we find that
$$N_+(t) = {1\over 2}[1- {\rm erf}(\bar x(t))]  \eqn\twoeleven$$
where
$$\bar x(t)= -{\bar q(t)\over \sqrt{\hbar \cosh 2t}}$$
and the error function is defined by
$$ {\rm erf}(z) = {2\over \sqrt\pi} \int_0^x dt\, e^{-t^2}$$
Note that with our choice of the classical trajectory \twosix-\twoseven,
$\bar x(t)$ is positive for all $t$.
It is easy to calculate the $t\to \pm \infty$ limits of \twoeleven:
$$\eqalign{
N_+(\pm \infty) =& {1\over 2}[1- {\rm erf}(\bar x(\pm \infty))] \cr
\bar x(\pm \infty)=& \sqrt{|E_0|\over \hbar}\exp(\mp\theta_0) \cr}
\eqn\twotwelve$$
where we have used the parametrization  \twoseven\
of $p_0, q_0$ in terms of $|E_0|,
\theta_0$.
For $\theta_0>0$, {\it i.e.} $p_0>0$, we see that
$N_+(\infty) > N_+(-\infty)$, showing that a finite amount
of ``trickling'' has taken place, the amount being
$$\t \equiv
N_+(\infty) - N_+(-\infty)= {1\over \sqrt\pi}
\int_{\bar x(-\infty)}^{\bar x(\infty)} dt \, e^{-t^2}
\eqn\twothirteen$$
Note that we find a positive ``trickle'' from the left to the right when
$p_0>0$. This is understandable because $p_0>0$ means that the mean
momentum of the wave-packet is also directed from the left to the right
(in the direction of increasing $q$).
For negative $\theta_0$, or equivalently negative
$p_0$, we find a negative value for $\t$, while $\t$ vanishes for
$\theta_0=0=p_0$.

\noindent For small $\theta_0$ we get
$$\t = 2 \theta_0\, \sqrt{|E_0|\over 2\pi\hbar} \exp[-{
|E_0| \over 2\hbar}] + o(\theta_0^2) \eqn\twofourteen$$
This can be compared with the leading WKB result for the tunnelling
amplitude, which is given by
$$\exp[-{1\over \hbar} \int_{-a}^a dx'\, \sqrt{V(x')- E_0}] \propto
\exp[-|E_0|\pi /\hbar \sqrt 2] \eqn\twofifteen $$
Here $\pm a$ are the classical turning points, satisfying $V(\pm a)=E_0$.

\section{\bf Time-dependent Classical Solution of $c=1$ String Field
Theory and $\exp(-C/\gst)$ Effects:}

In the last section we constructed an exact solution of the $u(p,q,t)$
theory which corresponds to  fermion number equal to one.  It describes
the phase space density of a single-fermion wave packet, part of which
tunnels through. To use this solution in constructing a solution of the
$N$-fermion problem we proceed as follows.
We consider the fermi sea of the
$N$-fermion system and imagine removing one fermion from the fermi level to a
`wave-packet  state' of the kind described above, with a mean energy that
is much higher than the fermi energy but still far lower than the top of
the potential barrier (we will presently make these statements more
exact).
The fermi sea of $N-1$ fermions corresponds to a phase space density
$$\eqalign{
u_0(p,q) =& \langle F| \int dx\; e^{-ipx}
\psi^\dagger (q-{\hbar x/2},t)
\psi(q+{\hbar x/2},t)  | F\rangle \cr
 =& \int^\mu d\nu\int dx\; e^{-ipx} \phi^*_\nu(q-{\hbar x/2})
 \phi_\nu(q + {\hbar x/2}) \cr
}
\eqn\threeone $$
Here $|F\rangle$ is the ground state of $(N-1)$ fermions,
$\phi_\nu(x)$  is
the eigenstate of the single-particle hamiltonian
$$\hat h= {1\over 2} ( {\hat p}^2- {\hat x}^2 + {g_3\over \sqrt N} {\hat
x}^3 + \cdots ) \eqn\threeonea $$
with energy $\nu$ and
$\mu$ is the fermi level for $(N-1)$ fermions.
In the limit $N\to \infty$, the right hand side of \threeone\ can be
evaluated explicitly. Denoting this limiting value by $\bar u_0(p,q)$, we
have
$$ \bar u_0(p,q) = {1\over 2\pi} \int_{-\infty}^\mu d\nu
\int_{-\infty}^\infty {d\lambda \over \cosh \lambda/2 } \exp i[\nu \lambda
- {1\over \hbar}(p^2 - q^2) \tanh {\lambda\over 2}]
\eqn\threeoneb $$
By construction, the $u_0(p,q)$ in \threeone\ satisfies the constraints
$$
\int {dp~dq \over 2\pi \hbar} u_0(p,q) = N-1
\eqn\threetwo
$$
$$
\cos {\hbar \over 2} \left(\partial_q \partial_{p'} - \partial_{q'}
\partial_p\right) u_0(p,q) u_0(p',q')\bigg|_{{p' = p \atop q' = q}} = u_0
(p,q)
\eqn\threethree
$$
The equation of motion is also satisfied since $u_0(p,q)$
is time independent and can be shown to depend on $p$ and $q$ only through
the classical hamiltonian $h(p,q) = {1\over 2} ( p^2 - q^2 + {g_3\over
\sqrt N} q^3 + \cdots)$.

There are two ways one can approach the problem of constructing the full
$u(p,q,t)$ that combines the phase space densities $u_0(p,q)$ and
$u_1(p,q,t)$. The first is to try to see if
$$ u(p,q,t) = u_0(p,q) + u_1(p,q,t)  \eqn\threefour$$
is a solution of the equations of motion and the constraints. It is easy
to see that in the large $N$ limit
the equation of motion and the total fermion number
constraint is satisfied since the corresponding equations are linear.
The quadratic constraint, however,
is not satisfied because of the cross term
\relax
$$C_{01}
\equiv \cos {\hbar \over 2} \left(\partial_q \partial_{p'} - \partial_{q'}
\partial_p\right) u_0(p,q) u_1(p',q',t)\bigg|_{{p' = p \atop q' = q}}
\eqn\threefive $$
Note, however, that in the classical limit $\hbar \to 0$,
$$ u_0(p,q) \to \theta (\mu - {p^2 - q^2\over 2}),\quad
{1\over 2\pi\hbar}
u_1(p,q,t) \to \delta (p- \bar p(t)) \delta (q- \bar q(t))
\eqn\threesix$$
where $\bar q(t)$ and $\bar p(t)$ are given by \twosix.
Clearly if in this limit we choose $(p_0, q_0)$, the initial ($t=0$)
position of the peak, to be
outside  the support of $u_0$, or alternatively the energy $E_0 \equiv
(p_0^2 - q_0^2)/2 $ to be greater than the fermi energy $\mu$, then
the cross term \threefive\ vanishes.
In other words,
$$ u(p,q,t) = \theta (\mu - {p^2 - q^2\over 2}) +
2\pi\hbar \delta (p- \bar p(t)) \delta (q- \bar q(t) \eqn\threeeight$$
is a solution of the equation of motion and the constraints in the limit
$\hbar \to 0$.

How about $\hbar \not= 0$? After all, if we put $\hbar=0$
the physical effect that we are after, the
``trickle'', vanishes. Let us assume that $\hbar$ is non-zero but
small. It is easy to see, by using  explicit expressions for $u_0(p,q)$ and
$u_1(p,q,t)$, that they develop exponential tails away from the support of
the $\theta$-function and $\delta$-function respectively. Thus, by
choosing the energy $E_0$ of the wave-packet to be  sufficiently far away
from the fermi energy $\mu$, we can make the cross term $C_{01}$ in
\threefive\ exponentially small. The region where the cross term is the
strongest is given by $(p,q) \approx (p_0, q_0)$ where $u_1$ is of order 1
and $u_0$ is of order $\exp[ - (a|\mu| - b\sqrt{|\mu E_0|})/ \hbar]$ ($a,
b$ positive numbers of order 1). If we choose $|\mu| >> |E_0| >>0$
then $C_{01} \sim \exp[ - a|\mu|/\hbar]$.
This implies that the solution $u(p,q,t) = u_0 + u_1$ is off from the
exact solution by terms of the order $\exp[ - (a|\mu|)/\hbar]$\foot{We
shall soon verify this statement explicitly by presenting the exact
solution.}. Note that we cannot outright ignore such terms  because the
``trickle'' that we are looking for is also exponentially small as $\hbar
\to 0$. The implication of this is the following. As in the previous
section,  let us  define the quantities $N_+(t)$ and $\t$ as
$$N_+(t) = \int_0^\infty dq\int_{-\infty}^\infty dp\; u(p,q,t)
\eqn\threenine$$
$$\t = N_+(\infty) - N_+(-\infty) \eqn\threeten $$
If $u= u_0 + u_1$ were the exact solution, then $\t$ would again be given
by \twothirteen-\twofourteen, since $u_0$ is time-independent and does not
contribute to the trickle. The question is, if $u$ has additional terms of
order $\exp[- (a|\mu|)/\hbar]$ (which are also clearly time-dependent)
then how does the estimate for the ``trickle'' modify? For instance, can
the new contribution cancel the trickle by contributing an equal
amount with an opposite sign?
Fortunately such bizarre things do not happen. The basic reason is that
expressions like \twothirteen\ or \twofourteen\ do not depend on $\mu$,
and in the
domain of parameters $|\mu| >> |E_0|>>0$ we can claim that, to leading
order, the trickle is again given by
$$\t = 2\theta_0\, \sqrt{|E_0|\over \pi\hbar} \exp[-{
|E_0| \over \hbar}] + o(\theta_0^2) \eqn\threeeleven$$

There is a more precise way of seeing the above result by going back to
fermions and constructing an exact classical solution $u(p,q,t)$ that
satisfies all the constraints and the equation of motion exactly. It is
given by
$$ u(p,q) = \langle \Phi| \int dx\; e^{-ipx}
\psi^\dagger (q-{\hbar x/2},t)
\psi(q+{\hbar x/2},t) | \Phi\rangle
\eqn\threetwelve $$
Here $|\Phi\ket $ is the $N$-fermion state in which $N-1$ fermions occupy
the $N-1$ lowest energy eigenstates $\phi_\nu(x),\, \nu=-M,\cdots, \mu$
($-M$ is a large negative number denoting the ground state energy of the
single-particle hamiltonian; in the limit of $N\to \infty$, $-M \to
-\infty$ and the energy levels become continuous) and one fermion belongs
to a wave-packet state, which is given, in the limit $N\to\infty$ in which
the single-particle hamiltonian is $\hat h={1\over 2}( {\hat p}^2 - {\hat
x}^2 )$, by
\relax
$$\eqalign{
\psi_1 & (x,t) = \exp({it\over 2}[\hbar\del_x^2 + x^2/\hbar])
\{(\pi \hbar)^{-1/4} \exp(-{1\over 2\hbar}[(x-q_0)^2 -2ip_0x])\} \cr
=&
{(\pi\hbar)^{-1/4}\exp(ip_0q_0/2\hbar)\over \sqrt{f(t)}}
\exp[-{1\over 2\hbar \cosh 2t}\{ (x f^*(t) - z_0)^2 +
{1\over 2}( |z_0|^2 \cosh 2t - z_0^2 f(t) )\}]
\cr
}
\eqn\threethirteen$$
where $z_0\equiv q_0 + ip_0$ and $f(t) = \cosh t + i \sinh t$.
$\psi_1(x,t)$ is the wave-function whose phase space density is
$u_1(p,q,t)$.
The state $|\Phi\ket$ is explicitly given by the Slater
determinant of the single-particle states $ \{\phi_\nu(x),$  $\nu=-M, \cdots,
\mu;$ $ \psi_1(x,t) \}$. To see that \threetwelve\ satisfies the quadratic
constraint \onesix\ it is convenient to define the first-order density
matrix [\DMWB] $\gamma_\Phi(x,y,t)= \bra \Phi| \psi^\dagger(x,t) \psi(y,t)
|\Phi \ket$ in terms of which the constraint  \onesix\ reads as
$\int dy\, \gamma_\Phi(x,y,t) \gamma_\Phi(y,z,t) = \gamma_\Phi(x,z,t)$.
The last statement is true for {\sl any} state $\Phi$ which can be written
as a single Slater determinant of any arbitrary one-particle
wavefunctions. Indeed, this is the easiest way to check the validity of the
quadratic constraint for $u_0$ and $u_1$ also.

Let us expand $\psi_1(x,t)$ in terms of the energy eigenfunctions
$\phi_\nu (x,t) \equiv \phi_\nu (x) \exp(-i\nu t)$:
$$ \psi_1(x,t)= \sum_\nu A(\nu) \phi_\nu(x) \exp(-i\nu t)
\eqn\threefourteen$$
Using these one can evaluate \threetwelve:
$$u(p,q,t) = \sum_{\nu({\rm sea})}  u_\nu(p,q) +  {1\over C^2} B
\eqn\threefifteen $$
where
$$ C^2= \sum_{\nu({\rm above})} | A(\nu)|^2 = 1- \sum_{\nu({\rm sea})}
|A(\nu)|^2 \eqn\threesixteen$$
and
$$ B= \sum_{\nu({\rm above})} |A(\nu)|^2 u_\nu(p,q)+
\sum_{\nu\not=\nu'({\rm above})} A^*(\nu) A(\nu') u_{\nu\nu'}(p,q,t)
\eqn\threeseventeen$$
In the above, sum over energy levels that belong to the fermi {\sl sea} or
{\sl above}  have been denoted appropriately. An unspecified sum means sum
over all states. The notation
$u_{\nu\nu'}(p,q,t) $ stands for
$\int dx\, \exp(-ipx)\phi^*_\nu(q-\hbar x/2,t)
\phi_{\nu'}(q + \hbar x/2,t)$.

It is clear that the first term in \threefifteen\ is simply
$u_0(p,q)$ of Eqn. \threeone. Let us compare the second term $B/C^2$ with
$u_1(p,q,t)$; the latter (Eqn. \twoone) looks in the present notation as
$$ u_1(p,q,t) =
\sum_\nu |A(\nu)|^2 u_\nu(p,q)+
\sum_{\nu\not=\nu'} A^*(\nu) A(\nu') u_{\nu\nu'}(p,q,t) \eqn\threeeighteen
$$
We see that, in the limit $N\to \infty$,  $B/C^2$ differs from
$u_1(p,q,t)$ by terms of the
order of $|A(\mu)|^2$. Now,
$A(\mu)$ is simply the scalar product between the
Gaussian wavefunction $\psi_1$ and the fermi level wave-function
$\phi_\mu$ and can be shown to be $\sim \exp[- a|\mu|/\hbar], a>0$
if we have $|\mu| >> |E_0| >>0$. This verifies our earlier conclusion that
$ u = u_0 + u_1 + o(\exp[-a |\mu|/\hbar]), \; a>0$.

\underbar{Stringy Non-perturbative effect:}

Since we are working in the weak coupling limit ($\hbar=\gst \to 0$), the
expression for the ``trickle'' that we have calculated using a classical
solution can be
regarded as the leading result for the following field theory amplitude
$$ {\cal A} \equiv {\int \D u \exp(iS[u]) \t[u] \over \int\D u \exp(iS[u])
}
\eqn\threenineteen $$
where
$$\t[u]=\int_0^\infty dq\int_{-\infty}^\infty dp [u(p,q,t=+\infty) -
u(p,q,t=-\infty)] \eqn\threetwo $$
In the above, $S[u]$ is the classical action described in Sec. 2 and the
measure $\D u$ incorporates the constraints on the $u(p,q,t)$ field. To
pick out the classical solution described above, we of course need to
specify boundary conditions in the functional integral appropriately
so that they match the behaviour of the desired classical solution
at large initial and final times.
By the results described above, we find that
$$ {\cal A} \sim \exp[-|E_0|/\gst]  \eqn\threetwoone$$
where $E_0 = (p_0^2-q_0^2)/2$ is a parameter of the classical solution
specifying the mean energy of the wave packet. As
already mentioned, the physics of this amplitude is the tunnelling of a
single fermion.

For treatment of non-perturbative effects within the framework of
collective field theory, see [\OVRUT-\JEVICKI]. Stringy non-perturbative
effects arising from the motion of a single eigenvalue in an effective
potential have been discussed previously in $c<1$ models in [\M].

\section{\bf Marinari-Parisi Model:}

In this section we briefly outline how the Marinari-Parisi model
[\MP]
can be
treated in our formalism of $u(p,q,t)$-theory so that non-perturbative
effects may be calculated in a field theory framework. The essential
point is that the bosonic sector of the model corresponds to a
non-relativistic fermi gas in one space dimension. The basic difference
with the $c=1$ model is that the classical single-particle  hamiltonian is
given by
$$h(p,q) = {p^2\over 2} + V(q),\quad V(q)=q^3-\alpha q \eqn\fourone$$
Thus, except for the equation of motion for the $u$-field, which becomes
$$ \del_t u = \{ h, u\}_{\sevenrm MB} =  \{ h, u\}_{\sevenrm PB} -
{\hbar^2\over 4} \del_p^3 u, \eqn\fourtwo $$
everything else (like the constraints, the classical action etc.) remains
unchanged.

The interesting physical effect in this model is assoicated with the
tunnelling of a single fermion. For $\alpha >0$, the potential  $V(q)$ has
two minima at $q=q_m\equiv +\sqrt{\alpha/3}$ and $q=-\infty$, separated by
a maximum at $q=-q_m$. It has been shown in [\MP] that at the critical
point $\alpha=0$, where the secondary well disappears, the ground state of
the fermi system is given by a fermi sea which reaches upto the position
of the point of inflexion. This means that as one decreases $\alpha$ from
the positive side towards zero, more and more fermions escape out of the
secondary well. In the limiting situation $\alpha \to 0^+$ only one
fermion remains and  criticality is characterized by the tunnelling of
this fermion.  This causes non-perturbative supersymmetry breaking,
leading to amplitudes that go as $\exp(- C/\gst)$. Since the tunnelling
involves a single fermion, the interpolating configuration $u(p,q,t)$
again consists of a ``large'' piece $u_0(p,q)$ describing $N-1$ stationary
fermions and a ``small'' piece $u_1(p,q,t)$.
We can explicitly construct $u_1(p,q,t)$
as follows.  Let $u_1$ at time $t=0$ be given by
$$ u_1(p,q,0)= 2
\exp[-{1\over \hbar}\{(p-p_0)^2+ (q-q_0)^2\}]. \eqn
\fourthree $$
It is easy to verify that \fourthree\ satisfies the constraints \twothree\
and \twofour\ at $t=0$. Now a useful fact about the time-evolution
$ \del_t u = \{ h, u\}_{\sevenrm MB} $ is that if we ensure that
$u(p,q,0)$ satisfies the two constraints \twothree\ and \twofour, then
the time-evolved $u(p,q,t)$ automatically satisfies them for any
{\sl arbitrary} hamiltonian $h(p,q)$ (this is of course required for
consistency between equations of motion and constraints). It is not
difficult to show that the solution to \fourtwo, with the initial
condition \fourthree, is
$$u_1(p,q,t) = 2 \exp[-{1\over \hbar}\{(P(p,q,t)-p_0)^2+
(Q(p,q,t)-q_0)^2\}] + o(\hbar^2). \eqn\fourfour$$
where $P(p,q,t), Q(p,q,t)$ describe a classical trajectory for the
hamilton \fourone, with the initial condition $P(p,q,0)=p, Q(p,q,0)=q$.

In order that \fourfour\ describes the appropriate tunnelling
configuration, we should take $q_0$ to be inside the secondary well ($q_0
\approx q_m\equiv \sqrt{\alpha/3}$) and $p_0$ to be negative such that the
wave-packet is directed towards the other well (the other well is strictly
speaking bottomless in the double scaled limit, but for any finite $N$
one has a regulated potential with a finite depth just as in the standard
$c=1$ model). The calculation of the ``trickle'' can again be performed in
a similar fashion to the earlier sections. We shall present the details
elsewhere.

\section{\bf Comparison with Collective Field Theory:}

In this section we ask whether our classical solution $u(p,q,t)$ could be
derived from the equations of motion of the standard
collective field theory
[\DJ]. The
answer will turn out to be negative. But before going to that, let us first
see how  one might make a comparison between the
$u(p,q,t)$-theory and the standard collective field theory.

It is convenient to define the following moments of the phase space
density $u(p,q,t)$:
$$\eqalign{
\rho(q,t) \equiv {\rhot(q,t)\over 2\pi\hbar}
= & \int {dp\over 2\pi\hbar} u(p,q,t) \cr
\Pi(q,t) \rho(q,t)=& \int {dp\over 2\pi\hbar} p u(p,q,t) \cr
\Pi_2(q,t) \rho(q,t)=& \int {dp\over 2\pi\hbar} p^2 u(p,q,t) \cr
 \cdots = & \cdots \cr}
\eqn\fiveone$$
In the following we shall assume that the $N\to\infty$ limit has been
taken. In this limit the classical hamiltonian is $h(p,q)= {1\over 2} (
p^2 - q^2) $ and therefore the equation of motion is
$$ (\del_t + p\del_q + q\del_p)u(p,q,t) =0 \eqn\fivetwo$$
This equation
implies equations of motion for the  moments. One can obtain them by
taking moments of \fivetwo. Let us write down the first two equations
obtained this way:
$$ \del_t \rhot(q,t) = -\del_q(\Pi \rhot)  \eqn\fivethree$$
$$ \del_t \Pi(q,t) = q + \del_q({\Pi^2\over 2} - \Pi_2)+ {\del_q \rho
\over \rho}( \Pi^2 - \Pi_2)  \eqn\fivefour$$
It is clear that one does not obtain a closed set of equations
for $\rho, \Pi$--- their equations of motion involve the next higher moment.
In fact this pattern continues {\sl ad infinitum}.

Let us compare \fivethree-\fivefour\
with  collective field theory equations
$$ \eqalign{
\del_t \rhot(q,t) =& -\del_q(\Pi \rhot) \cr
\del_t \Pi(q,t) =& q -\del_q({\Pi^2\over 2} + {\rhot^2\over 8}) \cr}
\eqn\fivefive $$
How does one understand getting a closed set of equations for $\rho, \Pi$
from our viewpoint? To see this, we turn again to quadratic profiles (for
details see [\DMWA, \DMWB])
for which the classical solution $u(p,q,t)$, in the
limit $\hbar \to 0$, looks like
$$u(p,q,t) = \theta[ (p_+(q,t)-p) ( p- p_-(q,t))]  +
\hbar\,{\rm corrections}
\eqn\fivesix $$
Remarkably, for these kinds of solutions we can show that the moment
$\Pi_2$ can be determined in terms of $\Pi, \rho$ as
$$ \Pi_2 = \Pi^2 + {1\over 12 } \rhot^2  + \hbar\,{\rm corrections}
\eqn\fiveseven$$
If one puts this in \fivefour, one recovers the second equation of
\fivefive\ upto $\hbar$-corrections (the first equation already agreed
with \fivethree).

There are two lessons to be learnt from the above exercise: (a) collective
field theory equations can be recovered from the equation of motion of
$u(p,q,t)$-theory under the assumption \fivesix, in the limit $\hbar \to
0$; (b) even under the ``quadratic profile'' assumption, {\bf classical}
equations of the collective field theory are violated by {\bf classical}
solutions of $u$-theory (like \fivesix) by $\hbar$-corrections. The last
observation reflects the fact that the classical solutions of the
$u$-theory incorporate the single-particle quantum mechanics exactly, as
was remarked in Sec. 3.
To give a more explicit example of this point, consider $u= u_0(p,q)$ of
the Sec. 4, and regard it for the moment as the fermi sea of the
$N$-body problem (rather than $N-1$). This classical solution satisfies
the ansatz \fivesix\ with non-trivial non-perturbative corrections in
$\hbar$. It is easy to see that the second equation of \fivefive\ is
violated by non-perturbative $\hbar$-corrections.

Let us now see if the full time-dependent $u(p,q,t)$ including $u_0$ and
$u_1$  of the previous section satisfies equations \fivefive. To a first
approximation, let us ignore the overlap regions between $(\rho_0,
\Pi_0 )$ and $(\rho_1, \Pi_1)$ and try to see if each of these pairs
satisfies the collective field equations independently. This amounts to
ignoring non-perturbative terms in $\hbar$ (recall that both $\rho_0$ and
$\rho_1$ have exponential tails in the intermediate region between them).
This attitude is similar to the one that we had adopted while calculating
the ``trickle''. In addition, since we have established non-perturbative
violations of the collective field equations already in the last
paragraph, one may be interested in looking for new violations this time
which persist even when one ignores terms of order
$\exp[ -1/\hbar]$. It turns out that $\rho_1, \Pi_1$, taken by themselves,
indeed violate the second equation of \fivefive\ by the amount
$$
\eqalign{
\del_t \Pi_1(q,t) -& \{q -\del_q({\Pi_1^2\over 2} + {\rhot_1^2\over
8})\}\cr
=& {q- \bar q(t) \over \cosh^2 2t}  -{\pi\over 2}{q- \bar q(t) \over
\cosh^2 2t}  \exp[- {2(q-\bar q(t))^2 \over \hbar \cosh 2t}] \cr}
\eqn\fiveeight$$
The non-perturbative term is already expected from earlier considerations;
its magnitude may change when one takes into account the overlap
terms between $\rho_0$ and $\rho_1$, though following the logic of
previous sections, the modifications are smaller than the original term.
The first term is more of a surprise because it is non-zero even in the
$\hbar \to 0$ limit. The way to understand this is to note that the
wave-packet solution $u_1(p,q,t)$ does not satisfy the criterion \fivesix\
appropriate for quadratic profiles. As a result it does not lead to
\fiveseven (which one may also verify directly). This is the reason why
there is a {\bf classical violation} of the collective field equations by
the wave packet solution.
Indeed, since the relation \fiveseven\ is rather crucial in deriving the
classical equations of collective field theory, and this  in turn
crucially depends on the assumption \fivesix\ of quadratic profiles, it
is trivial to generate other examples of $u(p,q,t)$ which in the limit
$\hbar \to 0$ go over to something other than quadratic profiles and thus
end up satisfying different collective field equations!

\section{\bf Interpretation of the Time-dependent solution in the Black
Hole Context:}

In [\DMWC] we found a  correspondence between the weak coupling regime of
$c=1$ string field theory and the black hole of two-dimensional string
theory.  One feature of the correspondence is that if one considers $\bra
R| u(p,q,t) | R\ket\equiv u_R(p,q,t) $ in states $| R\ket $ which are
`small fluctuations' on the ground state $|R_0\ket $ then the ``hyperbolic
transform''(HT)
of the fluctuation $u_R - u_{R_0}\equiv \eta $ satisfies, in
the classical limit, the differential equation of a massless scalar field
in black hole background
\foot{In equation (19) of [\DMWC] we made this claim for $\del_\mu \eta$.
By going through the steps that led to (19), we can show that the
equation is equally valid for $\eta$ itself.}. The `small fluctuation'
condition above means that the support of $\eta$ must be in a small
neighbourhood of the fermi surface, satisfying $|(h(p,q) - \mu) /
\mu | << 1$ whenever $\eta(p,q)\not= 0$.  Let us now see how we can find
such solutions from the ``tunnelling'' state that we have constructed and
used in the preceding sections to see somewhat different physical effects.

Consistent with the approximations that have been made in the earlier
sections, we shall regard the full solution for $u$ as
$$ u(p,q,t) = u_{N-1}(p,q) + u_1(p,q,t)  \eqn\sixone $$
where we have used the notation $u_{N-1}$ in place of $u_0$ to emphasize
that it corresponds to the fermi sea of an $(N-1)$-fermion system.
The fluctuation $\eta$ is the
difference between \sixone\ and the expectation value of the
$u(p,q,t)$-operator in the state $|R_0 \ket$ which describes the
$N$-particle fermi sea. We have
$$\eqalign{
\eta(p,q,t)
\equiv u(p,q,t)- u_N(p,q) &= -\delta_0 u(p,q) + u_1(p,q,t) \cr
\delta_0 u(p,q) = & u_N(p,q) - u_{N-1}(p,q) \cr
}
\eqn\sixtwo $$
$\delta_0 u(p,q)$ is simply the phase space density corresponding to the
fermion at the top of the $N$-particle fermi sea and $u_1(p,q,t)$ is the
phase space density of the wave-packet.
Now, by the arguments given in
[\DMWC], the HT (`hyperbolic transform') of $\delta_0 u(p,q)$, denoted by
$\delta_0 T(u,v)$, satisfies the differential equation
$$ [4(uv - {\mu\over 2}) \del_u \del_v + 2(u\del_u + v\del_v) + 1]
\delta_0 T(u,v) =0 + \hbar\,{\rm corrections} \eqn\sixthree $$
where $u,v$ are defined by $u={1\over 2}(p+q)e^{-t},\, v={1\over 2} (p-q)
e^t.$  This is because in the $\hbar \to 0$ limit the support of $\delta_0
u$ in the phase space is confined to a small strip near the fermi surface.
Now if we can ensure that in the $\hbar \to 0$ limit $u_1$ satisfies the
`small flucutation' condition mentioned in the last paragraph, then $\eta$
will also satisfy this condition and as a result the  HT of $\eta$ will
satisfy Eqn. \sixthree.  This would imply that the HT of $u_1(p,q,t)$,
would also satisfy Eqn. \sixthree. The `small
fluctuation' condition on $u_1(p,q,t)$ implies that we must have $|(E_0 -
\mu)/ \mu| << 1$.
If we go back to the arguments in Sec. 4, we can see
that in such a region of
parameters $p_0, q_0$, \sixone\ is a solution only in the classical limit
$\hbar \to 0$, since otherwise the cross terms discussed in \threefive\
are important. In the limit $\hbar \to 0$,
$$ u_1(p,q,t) \to 2\pi \hbar \delta (q- \bar q(t)) \delta (p - \bar p(t))
$$
The  HT of this limiting $\delta$-function is easy to calculate and turns
out to be proportional to
$$ T_1(u,v) = | (u-u_0) (v-v_0) |^{-1/2}
  \eqn\sixfour $$
where $u_0= {1\over 2}(p_0 + q_0)$ and $v_0 = {1\over 2} (p_0 - q_0)$.
It can be directly verfied that $T_1(u,v)$ satisfies the differential
equation \sixthree\ for $ |( u_0v_0 - \mu/2 )/\mu |<< 1$.

Note that in the limit that we are working with in this section, our
solution $u_1(p,q,t)$ does not exhibit any ``trickling'' and is not linked
with any non-perturbative effect. However, the solution $T_1(u,v)$ that it
gives rise to is rather interesting from the black hole point of view. Let
us emphasize that if one chooses parameters $u_0, v_0$ such that $u_0v_0 =
\mu/2$ then \sixfour\ provides an {\bf exact solution} of the
{\bf differential equation for }
{\bf propagation of massless scalar fields} {\bf in the
black hole geometry}:
$$ [4(uv - {\mu\over 2}) \del_u \del_v + 2(u\del_u +
v\del_v) + 1] T_1(u,v) =0  \eqn\sixfive $$
This solution has the intriguing feature that it has singularities along
the two lines $u=u_0$ and $v=v_0$. Since $\mu$ in our convention is negative,
$u_0v_0=\mu/2$ is satisfied by a family of values
$$ u_0=\alpha
\sqrt{-\mu/2},\; v_0 = -\alpha^{-1} \sqrt{-\mu/2} \eqn\sixsix $$
where $\alpha$ is a non-zero real number.

Let us try to understand this solution as a  tachyon wave in
the black hole geometry.
In the following we shall concentrate on the $v>0$ half of the Kruskal
diagram and
use ``space'' and ``time'' coordinates $\xi$ and $T$, related to $u,v$ by
$$ u = \epsilon \exp(\xi + T),\quad v= \exp (\xi - T) \eqn\sixseven $$
where $\epsilon = \pm 1$ depending on whether we are in the region $uv >0
$ or $uv <0$. Since $uv= \mu/2= -|\mu|/2$ denotes the position of the
black hole singularity in our convention, $uv>0$ or $\epsilon =+1$ denotes
spacetime regions outside the event horizon. The $\xi$ and $T$ coordinates
introduced above are simple functions of the Schwarzschild space and time
coordinates, respectively. The solution \sixfour\ then looks like
$$T_1(\xi, T) = |(\exp(\xi + T) -\epsilon \alpha \sqrt{|\mu|/2})
(\exp(\xi - T) +
\alpha^{-1} \sqrt{|\mu|/2} )|^{-1/2} \eqn\sixeight$$
Let us consider first the case $\alpha>0$. In this case, \sixeight\ has
singularities only outside the horizon $\epsilon >0$. At a fixed $\xi$
this singularity occurs at
$$ T = -\xi + {1\over 2} \log (|\mu|/2) \eqn\sixnine$$
This singularity can be interpreted in two ways. If we think of \sixeight\
as a propagating tachyon wave, then it implies that the initial data has a
singularity irrespective of the choice of the initial spacelike surface. A
more interesting interpretation comes about if we think of additional
observers who can couple to tachyon backgrounds. For such an observer at
fixed $\xi$, the tachyon solution \sixeight\ will appear as a singularity
at the instant of time $\sixnine$ and will be well-behaved before and
after. Since the lines of singularity of our solution $u=u_0$, $v=v_0$ are
light-like, these can perhaps be interpreted as light-like
``thunderbolts''
[\HAWKING]. The case $\alpha<0$ has the property that here one encounters
singularities only inside the event horizon. Note that the symmetry
between positive and negative values of $\alpha$ can be restored if one
includes in the discussion the other half of the Kruskal diagram $v <0$.
For further discussion of this and other solutions to
\sixfive, see [\DMWD].

\ni {\bf Acknowledgement:}  S.W. would like to thank S. Shenker and A.
Jevicki for discussions.

\refout
\end